\documentstyle[twoside,fleqn,espcrc2,epsfig]{article}
\newcommand{\AmS}{{\protect\the\textfont2
  A\kern-.1667em\lower.5ex\hbox{M}\kern-.125emS}}

\title{Chiral Symmetry Breaking in Strongly Coupled Quenched
QED$_4$ Using the Dyson-Schwinger Equation Formalism}
\author{
    A.G.\ Williams\address{
    Department of Physics and Mathematical Physics,
    University of Adelaide, Australia 5005}%
        \thanks{{\em e-mail:} awilliam@physics.adelaide.edu.au}
    and
    F.T.\ Hawes\address{Department of Physics B-159 and SCRI B-186,
    Florida State University, Tallahassee, FL 32306, USA}%
        \thanks{{\em e-mail:} hawes@mailer.scri.fsu.edu}
}

\begin{document}

\begin{abstract}
We study chiral symmetry breaking in quenched strong-coupling QED$_4$ in
arbitrary covariant gauge within the Dyson-Schwinger equation formalism.
A recently developed numerical renormalization program is fully implemented.
Results are compared for three different fermion-photon
proper vertex {\it Ans\"{a}tze\/}: bare $\gamma^\mu$, minimal Ball-Chiu,
and Curtis-Pennington.  The procedure is straightforward to implement
and numerically stable.
We discuss the chiral limit and observe that in this limit the renormalized
axial current is conserved.  A detailed study of residual gauge dependence
due to the vertex choice is in progress.  The relevance for lattice studies
is discussed. (hep-lat/9510011)
\end{abstract}

\maketitle

\section{INTRODUCTION}
Strong coupling QED in three space and one time dimension has been
studied within the Dyson-Schwinger Equation (DSE) formalism for some
time \cite{BJW,FGMS,Mandula}.  For a recent review of
Dyson-Schwinger equations and their application see for
example Ref.~\cite{TheReview}.
DCSB occurs when the fermion propagator develops a nonzero scalar
self-energy in the absence of an explicit chiral symmetry breaking
(ECSB) fermion mass.  We refer to coupling constants strong enough to induce
DCSB as supercritical and those weaker are called subcritical.
We write the fermion propagator as
\begin{equation}
  S(p) = \frac{Z(p^2)}{\not\!p - M(p^2)}
       = \frac{1}{A(p^2) \not\!p - B(p^2)}
\end{equation}
with $Z(p^2)$ the finite momentum-dependent fermion renormalization,
and $B(p^2)$ the scalar self-energy.
In the absence of an ECSB bare electron mass, by definition
DCSB occurs when $B(p^2)\neq 0$.
Note that $A(p^2)\equiv 1/Z(p^2)$ and $M(p^2)\equiv B(p^2)/A(p^2)$.

Many studies, even until quite recently, have used the
bare vertex as an {\it Ansatz\/} for the one-particle
irreducible (1-PI) vertex $\Gamma^\nu(k,p)$
despite the fact that this violates the Ward-Takahashi Identity (WTI)
\cite{WTI}.
With any of these {\it Ans\"{a}tze\/} the resulting fermion propagator
is not gauge-covariant, i.e., physical quantities such as the critical
coupling for dynamical symmetry breaking, or the mass itself, are
gauge-dependent \cite{KKM}.
A general form for $\Gamma^\nu(k,p)$ which does satisfy the Ward
Identity was given by Ball and Chiu in 1980 \cite{BC}; it consists of a
minimal longitudinally constrained term which satisfies the WTI, and a set
of tensors
spanning the subspace transverse to the photon momentum $q$.
Although the WTI is necessary for gauge-invariance, it is not a
sufficient condition; further, with many of these vertex
{\it Ans\"{a}tze\/} the fermion propagator DSE is not multiplicatively
renormalizable.  There has been much recent research on the use
of the transverse parts of the vertex to ensure both gauge-covariant
and multiplicatively renormalizable solutions
\cite{CPI,CPII,CPIII,CPIV,dongroberts,BashPenn},
some of which will be discussed below.

What was common to essentially all of the previous studies is
that the fermion propagator is not in practice subtractively renormalized.
Most of these studies have assumed an initially massless theory
and have renormalized at the ultraviolet cutoff of the integrations,
taking $Z_1 = Z_2 = 1$.
We describe here some results \cite{HW} of a study of subtractive
renormalization in the fermion DSE, in quenched strong-coupling
QED$_4$.  (In the context of this study of QED, the term ``quenched'' means
that the bare photon propagator is used in the fermion self-energy DSE,
so that $Z_3 = 1$.  Virtual fermion loops may still be present, however,
within the vertex corrections.)
Results are obtained for DSE with three different vertices:
the bare $\gamma^\mu$, the minimal Ball-Chiu vertex form \cite{BC},
and the Curtis-Pennington vertex \cite{CPI,CPII,CPIII,CPIV}.

\section{DSE and Vertex Ans\"{a}tze}
\label{sec_method}

The DSE for the renormalized fermion propagator, in a general covariant
gauge, is
\begin{eqnarray} \label{fermDSE_eq}
\lefteqn{S^{-1}(p^2) = Z_2(\mu,\Lambda)[\not\!p - m_0(\Lambda)] -
    i Z_1(\mu,\Lambda) e^2  }\nonumber \\
&& \times \int^{\Lambda} \frac{d^4k}{(2\pi)^4}
	  \gamma^{\mu} S(k) \Gamma^{\nu}(k,p) D_{\mu \nu}(q)\:;
\end{eqnarray}
here $q=k-p$ is the photon momentum, $\mu$ is the renormalization
point, and $\Lambda$ is a regularizing parameter (taken here to be an
ultraviolet momentum cutoff).  We write
$m_0(\Lambda)$ for the regularization-parameter dependent bare mass.
The physical charge is $e$ (as opposed to the bare charge $e_0$),
and the general form for the photon propagator is
\begin{displaymath}
  D^{\mu\nu}(q) = \left\{
    \left( -g^{\mu\nu} + \frac{q^\mu q^\nu}{q^2} \right)
    \frac{1}{1-\Pi(q^2)} - \xi \frac{q^\mu q^\nu}{q^2} \right\}
\end{displaymath}
with $\xi$ the covariant gauge parameter.  Since we will work in the
quenched approximation and the Landau gauge we have
$e^2 \equiv e_0^2 = 4\pi\alpha_0$ and
\begin{displaymath}
  D^{\mu\nu}(q) \to  D_0^{\mu\nu}(q)
    = \left( -g^{\mu\nu} + \frac{q^\mu q^\nu}{q^2} \right)
          \frac{1}{q^2}\:,
\end{displaymath}
for the photon propagator.

The requirement of gauge invariance in QED leads to the Ward-Takahashi
Identities (WTI); the WTI for the fermion-photon vertex is
$q_\mu \Gamma^\mu(k,p) = S^{-1}(k) - S^{-1}(p)$,
where $q = k - p$\/ \cite{BjD,MnS}.  This is a generalization of the
original differential Ward identity, which expresses the effect of
inserting a zero-momentum photon vertex into the fermion propagator,
$\partial S^{-1}(p)/\partial p_\nu = \Gamma^{\nu}(p,p)$.
In particular, it guarantees the equality of the propagator
and vertex renormalization constants, $Z_2 \equiv Z_1$.
The Ward-Takahashi Identity is easily shown to be satisfied
order-by-order in perturbation theory and can also be derived
nonperturbatively.

As discussed in \cite{TheReview}, this can be thought
of as just one of a set of six general requirements on the vertex:
(i) the vertex must satisfy the WTI; (ii) it should contain no kinematic
singularities; (iii) it should transform under charge conjugation ($C$),
parity inversion ($P$), and time reversal ($T$) in the same way
as the bare vertex, e.g., $C^{-1} \Gamma_\mu(k,p) C = -
\Gamma_\mu^{\sf T}(-p,-k)$
(where the superscript {\sf T} indicates the transpose);
(iv) it should reduce to the bare vertex in the weak-coupling
limit; (v) it should ensure multiplicative renormalizability of the
DSE in Eq. (\ref{fermDSE_eq});
(vi) the transverse part of the vertex should be specified to
ensure gauge-covariance of the DSE.

Ball and Chiu \cite{BC} have given a description of the most general
fermion-photon vertex that satisfies the WTI; it consists of a
longitudinally-constrained (i.e., ``Ball-Chiu'') part
$\Gamma^\mu_{\rm BC}$, which is a minimal solution of the WTI,
and a basis set of eight transverse vectors $T_i^\mu(k,p)$,
which span the hyperplane specified by $q_\mu T_i^\mu(k,p) = 0$,
$q \equiv k-p$.
The minimal longitudinally constrained part of the vertex is given by
\begin{eqnarray}
\lefteqn{\Gamma^\mu_{\rm BC}(k,p) = \frac{1}{2}[A(k^2) +A(p^2)] \gamma^\mu
+ \frac{(k+p)^\mu}{k^2-p^2} \times \nonumber}\\
&&\hspace{-0.7cm}\left\{ [A(k^2) - A(p^2)] \frac{{\not\!k}+ {\not\!p}}{2}
	      - [B(k^2) - B(p^2)] \right\}.
\label{minBCvert_eqn}
\end{eqnarray}
The transverse vectors can be found for example in Ref.~\cite{HW}.
A general vertex is then written as
\begin{equation} \label{anyfullG_eqn}
  \Gamma^\mu(k,p) = \Gamma_{BC}^\mu(k,p)
    + \sum_{i=1}^{8} \tau_i\,T_i^\mu(k,p)\:,
\end{equation}
where the $\tau_i(k^2,p^2,q^2)$ are functions which must be chosen to give the
correct $C$, $P$, and $T$ invariance properties.
Curtis and Pennington \cite{CPI,CPII,CPIII,CPIV} eliminate
four of the transverse vectors since they are Dirac-even and must
generate a scalar term.  By requiring that the vertex $\Gamma^\mu(k,p)$
reduce to the leading log result for $k \gg p$ they are led to
eliminate all the transverse basis vectors except $T_6^\mu$, with a
dynamic coefficient chosen to make the DSE multiplicatively
renormalizable.  This coefficient has the form
$\tau_6(k^2,p^2,q^2) = (1/2)[A(k^2) - A(p^2)] / d(k,p)$,
where $d(k,p)$ is a symmetric, singularity-free function of $k$ and $p$,
with the limiting behavior $\lim_{k^2 \gg p^2} d(k,p) = k^2$.
[Here, $A(p^2)\equiv 1/Z(p^2)$ is their $1/{\cal F}(p^2)$.]
For purely massless QED, they find a suitable form,
$d(k,p) = (k^2 - p^2)^2/(k^2+p^2)$.  This is generalized to the
case with a dynamical mass $M(p^2)$, to give
$d(k,p) = [(k^2 - p^2)^2 + [M^2(k^2) + M^2(p^2)]^2]/(k^2+p^2)$.

\section{The Subtractive Renormalization}
\label{sec_subtr}

As discussed in Ref.~\cite{TheReview} one first determines a finite,
{\it regularized\/} self-energy,
which depends on both a regularization parameter and the
renormalization point; then one performs a subtraction at the renormalization
point, in order to define the renormalization parameters $Z_1$, $Z_2$, $Z_3$.
The DSE for the renormalized fermion propagator,
\begin{eqnarray} \label{ren_DSE}
  \widetilde{S}^{-1}(p) & = & Z_2(\mu,\Lambda) [\not\!p - m_0(\Lambda)]
    - \Sigma'(\mu,\Lambda; p) \nonumber\\
    & = & \not\!p - m(\mu) - \widetilde{\Sigma}(\mu;p)\:,
\end{eqnarray}
where the (regularized) self-energy is
\begin{equation} \label{reg_Sigma}
  \Sigma'(\mu,\Lambda; p) = i Z_1 e^2 \int^{\Lambda}
    \frac{d^4k}{(2\pi)^4} \gamma^\lambda \widetilde{S}
      \widetilde{\Gamma}^\nu
      \widetilde{D}_{\lambda \nu}\:.
\end{equation}
[To avoid confusion we will follow Ref.~\cite{TheReview} and in this
section {\it only} we will denote
regularized quantities with a prime
and renormalized ones with a tilde, e.g. $\Sigma'(\mu,\Lambda; p)$
is the regularized self-energy depending on both the renormalization
point $\mu$ and regularization parameter $\Lambda$
and $\widetilde{\Sigma}(\mu;p)$ is the renormalized self-energy.]
The self-energies are decomposed into Dirac and scalar parts,
$\Sigma'(\mu,\Lambda; p) = \Sigma'_d(\mu,\Lambda; p^2) {\not\!p}
+ \Sigma'_s(\mu,\Lambda; p^2)$
(and similarly for the renormalized quantity,
$\widetilde{\Sigma}(\mu,p)$).
By imposing the renormalization boundary condition,
$\left. \widetilde{S}^{-1}(p) \right|_{p^2 = \mu^2} = {\not\!p} - m(\mu)$,
one gets the relations
$\widetilde{\Sigma}_{d,s}(\mu; p^2) =
\Sigma'_{d,s}(\mu,\Lambda; p^2) - \Sigma'_{d,s}(\mu,\Lambda; \mu^2)$
for the self-energy,  $Z_2(\mu,\Lambda) = 1 + \Sigma'_d(\mu,\Lambda; \mu^2)$
for the renormalization, and $m_0(\Lambda) =
\left[ m(\mu) - \Sigma'_s(\mu,\Lambda; \mu^2) \right]/ Z_2(\mu,\Lambda)$
for the bare mass.
The chiral limit is approached by taking
$m(\mu) = \Sigma'_s(\mu,\Lambda; \mu^2) = 0$ for fixed $\mu$.  This is
the limit in which the renormalized axial vector (non-flavor
singlet) current is conserved.
As $\Lambda\to\infty$ the numerical results are consistent with
the bare mass also vanishing as one would expect.  The mass renormalization
constant is given by $Z_m(\mu,\Lambda) = m_0(\Lambda)/m(\mu)$.
The vertex renormalization, $Z_1(\mu,\Lambda)$ is identical to
$Z_2(\mu,\Lambda)$ as long as the vertex {\it Ansatz\/} satisfies
the Ward Identity; this is how it is recovered for multiplication
into $\Sigma'(\mu,\Lambda;p)$ in Eq. (\ref{reg_Sigma}).
Since for the bare vertex case there is no way to
determine $Z_1(\mu,\Lambda)$ independently we will also in this
case use $Z_1 = Z_2$ for the
sake of comparison.

\section{Results}

\begin{figure}[htb]
\vspace{-7pt}
\setlength{\epsfxsize}{7.3cm}
\rightline{\epsffile{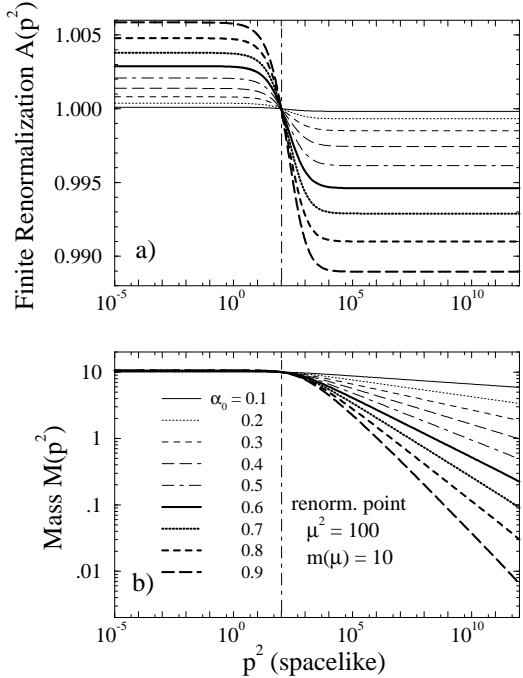}}
\vspace{-36pt}
\caption{Family of DSE solutions with the
    Curtis-Pennington vertex and subcritical couplings $\alpha_0 = 0.1$
    to 0.9 in the Landau gauge.  The renormalization point is
    $\mu^2 = 100$, and
    renormalized mass is $m(\mu)=10$ for all cases shown.
    Ultraviolet cutoffs are $1 \times 10^{12}$ for all cases.
    (a) the finite renormalizations $A(p^2)$;
    (b) the mass functions $M(p^2)$.}
\vspace{-12pt}
\label{fig_1}
\end{figure}

Solutions were obtained for the DSE with the Curtis-Pennington and
bare vertices, for couplings $\alpha_0$ from 0.1 to 1.75;
solutions were also obtained for the minimal Ball-Chiu vertex, with
couplings $\alpha_0$ from 0.1 to 0.6 (for larger couplings the DSE with
this vertex was susceptible to numerical noise).
For the solutions in the subcritical range, the renormalization point
$\mu^2 = 100$, and renormalized masses were either $m(\mu) = 10$ or
30.  Ultraviolet cutoffs were $1.0\times 10^{12}$ and
$1.0\times10^{18}$.
The family of solutions for Landau gauge and the Curtis-Pennington vertex
with $m(\mu) = 10$ is shown in Fig.~\ref{fig_1}.
Fig.~\ref{fig_1} a) shows the finite renormalization
$A(p^2)$; note that for the solutions with bare $\gamma^\nu$, we
would have $A(p^2) \equiv 1$ for all values of the coupling.
Fig.~\ref{fig_1} b) shows the masses $M(p^2)$.
Since the equations have no inherent mass-scale, the cutoff
$\Lambda$, renormalization point $\mu$, $m(\mu)$, and units of
$M(p^2)$ or $B(p^2)$ all scale multiplicatively, and the units are
arbitrary.
The results for all vertices are qualitatively similar.
For the Landau gauge results we find that as
the value of $\mu$ increases, $Z_2$
approaches 1, and also that $Z_2$ is almost independent of the
cutoff $\Lambda$, at least for $\Lambda \gg \mu$.
The solutions are extremely stable as the cutoff is varied.

In all cases the mass and finite renormalization were stable with respect
to very large variations in cutoff.
The renormalization constants $Z_2(\mu,\Lambda)=Z_1(\mu,\Lambda)$ remain
finite and well-behaved with increasing $\Lambda$ in contrast to
what happens in perturbation theory.

\section{Summary and Conclusions}

A full discussion of the Landau gauge results can be found in Ref.~\cite{HW}.
Results have also been recently obtained in other covariant gauges
and this work is currently being prepared for publication.
A careful comparison with lattice formulations of gauge-fixed QED can
now be attempted, since we are free to choose the renormalization point in
this work as the zero-momentum point for the $n$-point Green's functions
just as is conventionally done in lattice gauge theory studies.  Such
comparative studies should yield significant benefits.

This work was partially supported by the Australian Research Council,
the U.S. Department of Energy, the Florida State University
Supercomputer Computations Research Institute, and by grants of
supercomputer time from the U.S. National Energy Research Supercomputer
Center and the Australian National University
Supercomputer Facility.

%=======================================================================
%          Bibliography:
%-----------------------------------------------------------------------

\end{document}